\documentclass[12pt]{article}

\begin{document}

\author{Dario Sassi Thober \and Center for Research and Technology CPTec - Unisal \and thober@fnal.gov thober@cptec.br \and Phone.55-19-7443107 Fax.55-19-7443036}
\title{The definition of a magnetic monopole in Electrodynamics combined with
Gravitation}
\date{}
\maketitle

\begin{abstract}
It is discussed the singular string associated to the gauge field of
monopoles must be a physical observable if the monopole charge $g$ is
different from zero, $g\neq 0$. It is naturally found that if the gauge is
to be an observable, it is possibly connected to gravity.
\end{abstract}

\section{Introduction}

Some non solved fundamental theoretical questions are involved in the
definition of the magnetic charge or monopole in Quantum Electrodynamics.
One of these questions is the non - observable singular gauge string (many
times called Dirac string). Electric charges must never touch such strings
according to the original Dirac's proposal, [1,2]. In section 2 it is
discussed the general problems related to Dirac monopoles [3,4] and the
topologycal solution of Wu and Yang [5,6,7] where the gauge theory is
identified to a Fiber - Bundle geometric structure. According to this view
Dirac strings are said to be non - observable. In the same section we cite
the important recent work of He, Qiu and Tze [8] in which it is proven no
magnetic charge different from zero in Quantum Electrodynamics is possible
if the string is not a physical observable. This cleary shows that pure
Quantum Electrodynamics (as we know it) does not supports monopoles in any
way.

We follow identifying that in general the monopole string may have some
volume in the three - dimensional flat world. This implies that for $g\neq 0$
(the monopole's charge different from zero) one has to consider a reduced
physical world for the particles and fields in interaction with the
monopole. The string is the place where the gauge potential is not defined,
and as a consequence (in the general case of volumetric string) the
monopole's magnetic field is also not defined in this region. In section 3 we consider Gravity as
seen in a three - dimensional flat world, showing there are forbidden places
for particles and fields in the case even a small spherical mass is present.
Defining the electromagnetic vector potential as having a volumetric
singular string, it is the same as to define it in a reduced physical space
relative to the three - dimensional flat space. As the physical spacetime in
the presence of some mass can be seen as this smaller or reduced space
(subspace in some formulations, [6]), the monopole is defined in the
combined theory of Electromagnetism and Gravity. This means monopole gauge strings
are observable as spacetime distortions. Other interpretations and studies
on this connection are addressed to a future work in the concluding
subsection of section 3.

\section{The definition of a monopole}

Magnetic charge or monopole is a open theoretical and experimental question.
The original version of the monopole proposed by Dirac [1] never found a
place in Nature as we know it: It has never been found by any tangible
experiment and the theoretical machinery used to define its properties never
fulfil all requirements in terms of what we understand by a particle in
Electrodynamics.

The string of singularities has been a problem of non - natural assumptions
in Quantum Electrodynamics since Dirac's proposal [1]. In order to avoid the
string Dirac defined a {\it{nodal line}}, a region of the spacetime where
all Schroedinger wave functions associated to the particles in the Universe
are identically zero. In the Schroedinger equation, $(-i\mathbf{\nabla }-e%
\mathbf{A})^{2}\psi =2m\mathcal{E}\psi ,$ it is necessary for $\psi $ to be
zero or discontinous over the string of singularities. The nodal line is
then a constraint on the places the particles can be found - no particle is
allowed to be in some determined (by the strings positions) places of the
Universe.

After some time of theoretical existence, the monopole idea was improved by
Dirac himself in terms of a variational principle formulation and the
problems related to it, [2]. The source of problems again came related to
the definition of the singularity line of the vector potential. It is
necessary to assume that a string of such type never pass through a charged
particle if the equations of motion are to be derived from an action
principle. In trying a quantized version of the theory via a Hamiltonian
formulation, Dirac [2] found that the Poisson Brackets for the vector -
potential field is directly dependent on an arbitrary function which, in its
turn, is related to the mechanism used to define a possible action principle
(i.e., without the divergences that arise from the singular strings). Up to
that point the string was the theoretical difficulty which required
something the basic principles in Electrodynamics could not give: A
topologycal explanation in which the string could find a place as a gauge
artifact.

Almost twenty years later, Rohrlich [3] and Rosenbaum [4] showed in a
different approach the conditions for the existence of a variational
principle valid in the case of monopoles. It is interesting to quote the
problem put forward by Rohrlich that a system of particle equations can be
derived only from a nonlocal action integral and that no action integral
exists from which both the particle and field equations can be derived. As
in Rosenbaum work, the idea is to have a non - natural contraint (not
derived from the action principle) about the dynamics of charges and
monopoles: Charges must never touch monopoles; Since Lorentz force can be
derived from a principle which states that a charge approaching a monopole
along a straight line will collide, it is in contradiction with the
necessity of a charge never pass through a monopole. These considerations
indicated that something strange to Electrodynamics should be assumed in
order to not fall into contradiction about the theoretical existence of the
monopole. In fact, some years later t 'Hooft and Polyakov [5] proposed that
a natural condition for the monopole to exist will arise when other topologies
are involved, i.e., when other forces (associated to broken non - Abelian
gauge theories) of Nature are in consideration.

About the same time Wu and Yang [6] gave a clever topologycal description of
the problem in non - Abelian as well as in the Abelian case. Using the
concept of Fiber Bundles their proposal is that along a path of some charged
particle (in the $U(1)$ case), the gauge can be changed in order to avoid
the string of singularities. Each gauge changing is well defined if Dirac
Quantization Condition is respected - which is quite obvious since a phase
changing of $2\pi $ will not alter Physics. In this view no singularities
are seen by any particle's path and a beautifull connection to geometric
concept of fields is naturally given.

Soon after Wu and Yang's work, Brandt and Primack [7] showed that original
Dirac's theoretical formulation [1] of the monopole is equivalent to the Wu
and Yang's one [6]. Showing that the Dirac string attached to the monopole
can be arbitrarily moved by a gauge transformation valid everywhere and that
it can be completely removed at the cost of introducing a non - global
topology in space, Brandt and Primack accomplished to demonstrate the
equivalence of the formulations.

Recently He, Qiu and Tze [8] proposed the inconsistency of Quantum
Electrodynamics in the presence of monopoles. These authors [8] formally
show that the gauge coupling associated with the unphysical longitudinal
photon field is non - observable and has an arbitrary value in Quantum
Electrodynamics. In deriving Dirac Quantization Condition in $U(1)$ case it
was found that this derivation involves only the unphysical longitudinal
coupling constant. This work was focused on Dirac point - monopoles in the
standart Dirac formulation [1,2]. It is interesting to observe that
according to Brandt - Primack [7] and He - Qiu - Tze [8] works, Quantum
Electrodynamics is inconsistent in the presence of monopoles in whichever
formulation. The singular Dirac string in the monopole gauge potential is a
purely gauge artifact, it is just a gauge freedom which allows one to
arbitrarily move the string around without any physical effect, provided
Dirac Quantization Condition is satisfied. He, Qiu and Tze [8] observed that
by introducing another unphysical pure gauge field into Quantum
Electrodynamics, it is possible to attribute part of the singularities to
this pure - gauge field, and thus the corresponding Quantization Condition
involves the unphysical gauge coupling associated with this pure - gauge
field. After Dirac monopoles are introduced in whichever way, the exact $U(1)
$ gauge invariance must be respected, so that fixing any specific physical value
for an unphysical gauge coupling will violate the exact $U(1)$ gauge
invariance. Conversely, {\it{if we assume}} $g\neq 0$ \textit{the string
must be observable }according to reference\textit{\ }[8]. It is impossible
to have $g\neq 0$ and require no physical effects due to the singular string
as Dirac [1,2] or Wu and Yang [6] claimed, [8]. This implies that some new
fundamental principle should exist, which gives this pure - gauge coupling
(and consequently to the longitudinal photon field) a physical meaning so
that it become observable [8]. Accordingly, He, Qiu and Tze thought this is
most unlikely and set $g=0$ as the only possible solution in pure Quantum
Electrodynamics.

Aharonov and Bohm proposed an experimental test for the quantization of the
magnetic flux related to the discret value of the electron's charge, [9]. A
magnetized iron filament called whisker is positioned inside the volume of
the environment where electrons are allowed to perform trajectories, i.e.,
these electric charges are not allowed to go inside the whisker region. This
simple case has no problem to be defined theoretically: The reason is
because there is a physical limitation (some material in the whisker region)
for the electric charges to not cross over the magnetized filament.

It turns out to be impossible to define a singularity - free potential $%
A^{\mu }$ over all the flat three - dimensional world (i.e., over all $%
\bf{R}^{3},$[6]). In fact as in reference [6] any fiber of the physical
environment is smaller than $\bf{R}^{3}$, the physical space cannot be
defined as the entire $\bf{R}^{3}$ flat world for a given gauge.

The problem related to the monopole's string can be understood in terms
of the singular region associated to the electromagnetic potential $A^{\mu }$%
, without which no quantization of the angular momentum $eg/c$ is defined ($%
e $ is the quantum of electric charge and $g$ is the quantum of magnetic
charge). The main problem is that this singular part of the $A^{\mu }$
potential implies an infinite amount of interaction energy (self or mutual).
A simple example of a vector potential is: 
\begin{equation}
{\bf{A}}=-g\frac{1+\cos \theta }{r\sin \theta step\left[ \theta
-\delta \right] }{\bf{\phi }}
\end{equation}

where $\bf{\phi }$ is the unitary vector associated to the spherical
coordinate $\phi ,$ and $\theta $ is the other spherical coordinate in $%
\bf{r}=\left( r,\theta ,\phi \right) $ defined in a \textit{three -
dimensional flat spacetime} ($\bf{R}^{3}$ according to the terminology
of this work). The ``step'' function is defined as $step
[x-x_{0}]=0(x<x_{0});1/2(x=x_{0});1(x>x_{0})$, and $\delta $ is a vanishing
angle, $\delta \rightarrow 0$. The magnetic field derived from this
potential ($\bf{B=\nabla }\times \bf{A})$ is: 
\begin{equation}
{\bf{B}}=g2\pi (1+\cos \delta )\frac{\bf{r}}{r^{3}}
\end{equation}

if $\delta <\theta \leq \pi $, any $\phi $, otherwise the magnetic field is
not defined. The energy proportional to $\int \bf{B}_{i}\cdot \bf{B}%
_{j}d^{3}x$ (the indices refer to different sources $i\neq j$, or the same
source $i=j$) is not defined for $\theta \leq \delta $, any $\phi $ (the
string region), as well as the momentum proportional to $\int \bf{E}%
_{i}\times \bf{B}_{j}d^{3}x$.

When $\delta =0$ it is said that $\bf{A}$ remains singular but $\bf{B%
}$ is well defined everywhere. The general situation happens when the
singular string has some volume however ($\delta \neq 0$ in the example).

The theoretical question about Dirac monopoles can be described in two
items: \textit{i}) The general formulation in terms of an action principle
which allows an Hamiltonian formulation and consequent quantization, and 
\textit{ii}) The problem involved in defining a hidden (non observable)
string of singularities. The main pourpose of the present work is to deal
with the string problem (\textit{ii}) and address to the first (\textit{i})
in another work.

Lets consider the magnetic field is given by the rotational of the vector
potential, $\bf{B=\nabla }\times \bf{A}$, in the example as given in
equations (1) and (2). If it is said $g\neq 0$ then it must be assumed the
physical space where the magnetic flux is measured, $\oint \bf{B\cdot
d\sigma }$ ($\bf{d\sigma }$ the elementary oriented area surface), is
smaller than the $\bf{R}^{3}$. \textit{In saying this flux has a non -
zero net value, one is assuming the space is no longer the entire three -
dimensional flat because nor particles or fields are allowed to be in string
region}. The direction of the singular string in $\bf{R}^{3}$ world has
no physical meaning since now it is considered no hole exists in the \textit{%
reduced space} (the actual physical space where $g\neq 0$).

The strings have some three - dimensional volume in general. The general
case for bosons and fermions particles studied by Weisskopf [10] put
limitations on this particles radii even in the general quantum -
relativistic formulation. Of course the radius of some particle is related
to the not - well defined interaction between two points which are the same
in spacetime. This issue is a specific part of the definition of particles
which is not the focus in the present work, and it will be considered in
general that a string which is to be attached to a particle must have some
nonzero volume accordingly, [10]. Observe that the problem Dirac [1,2] faced
with the strings is of the same kind Weisskopf did, [10]: The fields must
(in Dirac's words, [2]) \textit{go out of existence} at some places in order
to define particles. The news in the monopole case is that this is an
extended region in space.

In the next section we intend to provide a definition of $g\neq 0$ in the
combined situation of having Electromagnetism and Gravitation. This
conjunction provides a natural place for the singular string of the gauge
fields. This formulation is different from that of Dirac [1,2] or Wu - Yang
[6] because the effect of the string is a physical observable.

\section{The observable singular string}

In the last section it was discussed that the existence of a monopole in $%
U(1)$ case ($g\neq 0$ in Quantum Electrodynamics) implies in general the
space ones takes as physical is in fact smaller than that known as the flat $%
\bf{R}^{3}$. The direction of the undefined region in $\bf{R}^{3}$
has no physical meaning once one accepts $g\neq 0$. It is very interesting
and important to observe that this comes from the fact $\bf{A}$ \textit{%
is not defined} in the string region - the value $1/0$ (one over zero) has
no physical or mathematical meaning in describing fields.

In order to consider $g\neq 0$ it is necessary to have a physical reason to
the fields and particles never be defined in the string region. We found
gravity is a possible candidate to hide the string.

\subsection{A physical place for the singular string}

Now it is considered the effects of Einstein's gravity in terms of
deformations in spacetime as viewed in the flat three - dimensional world.
This is a known valid and feasible way to see gravity, [11].

The curved spacetime interval outside the region of some matter distribution
can be written as: 
\begin{equation}
ds^{2}=A(r)dt^{2}-B(r)dr^{2}-C(r)r^{2}d\theta ^{2}-C(r)r^{2}\sin ^{2}\theta
d\phi ^{2}
\end{equation}

where $r$, $\theta $, $\phi $ are regarded as spherical coordinates and $%
A(r) $, $B(r)$, $C(r)$ are given functions of $r$. It is possible to show
that for $r=R$ the corresponding physical area $\Delta $ for $R$ fixed is: 
\begin{equation}
{\Delta }=4\pi R^{2}C(R)
\end{equation}

and that the physical distance $\Lambda $ between the points $r=R_{0}$ and $%
r=R$ on a given radial line is: 
\begin{equation}
\Lambda =\int_{R_{0}}^{R}\sqrt{B(r)}dr.
\end{equation}

Consider a spherical mass $m$ at the center of the coordinate system with
some given radial matter distribution function. In this particular case the
interval for $r>2m$ is: 
\begin{equation}
ds^{2}=(1-\frac{2m}{r})dt^{2}-(1-\frac{2m}{r})^{-1}dr^{2}-r^{2}d\theta
^{2}-r^{2}\sin ^{2}\theta d\phi ^{2}
\end{equation}

according to Schwarzschild. In this particular case the function $C(r)$ is
equal to one. We are interested in to study this metric as seen in the $%
\bf{R}^{3}$ world (three - dimensional flat). For the parameter $r=R,$
the physical area ${\Delta }$ is $4\pi R^{2}.$ Let us determine the
physical radius $\Lambda $: 
\begin{equation}
\Lambda =\int_{R_{0}}^{R}r/\sqrt{r^{2}-2mr}dr=\left[ \sqrt{\left(
r^{2}-2mr\right) }+m\ln \left( r-m+\sqrt{\left( r^{2}-2mr\right) }\right)
\right] \allowbreak _{R_{0}}^{R},
\end{equation}

where $R_{0}$ is some internal radius ($R_{0}<R$) in the region of the
matter distribution. We assume that the resulting value of the expression
above for the constant parameter $R_{0}$ is vanishingly small (of the order
of $2m$) compared to the one for $R$, so that the physical radius is: 
\begin{equation}
{\Lambda }=\sqrt{\left( R^{2}-2mR\right) }+m\ln \left( R-m+\sqrt{%
\left( R^{2}-2mR\right) }\right)
\end{equation}

and for $R>>2m$, 
\begin{equation}
{\Lambda }=R+m\ln \left( 2R-2m\right)
\end{equation}

i.e., the physical distance ${\Lambda }$ is larger than the parameter 
$R$.

The conclusion is that for some physical distance ${\Lambda }$ the
area avaliable to cover a sphere of this radius is $4\pi R^{2}$, with $R<{\Lambda }$, in the $\bf{R}^{3}$ world. In the three -
dimensional flat world it is impossible to close the surface of radius $%
{\Lambda }$ in this case. For a particle in the physical avaliable
world no hole occurs, it is all continuous, but in the $\bf{R}^{3}$
world there is a region where the spacetime is not defined. As each
spherical surface has this non - physical region, it performs a volume of
forbidden places for all particles and fields in $\bf{R}^{3}$.

It is known that for a flat three - dimensional world it is possible to
define an average curvature by means of a defect from $4\pi r^{2}$ of the
measured area of some surface of radius $r$. The connection of this idea to
the theory of gravitation is via a conceptual significance of the $G_{4}^{4}$
component of the stress - energy tensor. It is the average curvature $%
R_{12}^{12}+R_{23}^{23}+R_{13}^{13}$ of the three - space, which is
perpendicular to the time. This is a known valid interpretation of the
theory of gravitation, [11].

Now we have two physical results in $\bf{R}^{3}$: The electromagnetic
potential $\bf{A}$ according to Maxwell's Electromagnetic theory and the
physical region for particles and fields according to Einstein's
gravitational theory. The electromagnetic potential is physical only outside
the singular (in general volumetric) string region. The spacetime is only
physical outside some region defined by the holes on each spherical surface
starting from the monopole of mass $m$, as viewed in the flat $\bf{R}%
^{3} $.

\textit{The two forbidden regions in the }$\bf{R}^{3}$\textit{\ world
can be set to be the same}. It is only necessary to define a volumetric
string as source for the electromagnetic potential (where $\bf{A}$ is
not defined) exactly in the region the physical space is also not defined in
the flat world. The measure $\oint \bf{A\cdot dl}$ can be set to be a
constant for any closed loop around the string (in the physical region
already) if we require a constant flux at the string for any distance from
the monopole.

This only means that in order to define a monopole the spacetime must be modified accordingly. The same situation happened to Weisskopf [10], who defined a forbidden spherical region in spacetime where {\textit{all}} the fields are non - existent. In the monopole case this forbidden region is extended in spacetime.

\subsection{ A fiber - bundle formulation}

It is possible to describe the resulting physical situation of a reduced
space due to gravitation in a fiber - bundle formalism when we identify the
gauge given by $\bf{A}$ in $\bf{R}^{3}$ as determining the region of
the singular string (or the forbidden region in $\bf{R}^{3}$). As in Wu
- Yang formalism [6], each gauge determines a fiber or subspace which is
smaller than the flat three - dimensional world.

The view in Wu and Yang's work is that by the gauge freedom it is possible
for a particle to be defined in all $\bf{R}^{3}$ since it is possible to
rotate some subspace (which is smaller than the entire $\bf{R}^{3}$) to
turn it into another, making some forbidden region avaliable.

In interpreting the effect of gravity as a reduction of the physical space
regarded to the flat three - dimensional world, there exists a forbidden
region for particles in this world. This region can be arbitrarily set on
some direction in the referred space with no physical consequences
whatsoever. We defined a gauge where the singular string is at the same
region in $\bf{R}^{3}$ where particles are forbidden to go due to
gravity, so that a gauge is associated to the position of the string in flat
three - dimensional space.

In any of the two descriptions (the present one or Wu - Yang's) each gauge
is associated to a reduced space relative to the flat three - dimensional
one. The difference between the two formalisms comes from the fact that in
the present case the interpretation is so that the reduced space for some
gauge means in fact the physical avaliable space is reduced, i.e., the
spacetime is distorted (and strings are observable); Changings on the string
position will not result all $\bf{R}^{3}$ can be visited.

\subsection{The fundamental question and conclusions}

The fundamental question about a non - physically observable gauge string is
that $g$ must be exactly null if gauge invariance is to be respected in
Quantum Electrodynamics. This is one of the main causes all theoretical
formulations about Dirac monopoles are inconsistent [8].

The present work proposes a definition of the monopole in a combined
situation where Gravity and Electrodynamics are considered at the same time.
It is discussed the electric charges and fields never touch a monopole
string due to Gravity, i.e., Gravity becomes the physical observable if $%
g\neq 0$. It is in agreement with the conclusions in reference [8] where it
is shown $g\neq 0$ implies the string must be a physical observable.

He, Qiu and Tze, [8], discussed also that in the case of a point monopole
the physical observable is connected to the longitudinal photon field. In
ordinary Quantum Electrodynamics this field is not physical and consequently 
$g$ must be zero in pure QED (because the longitudynal photon field is not a
physical observable). In view of the present result we can confirm that
string is physical when $g\neq 0$ in the general case when it has
some volume in $\bf{R}^{3}$, but it is not clear if in this case the
longitudinal photon field has some direct role. This deserves more
investigation in a future work.

\section{Acknowledgements}
I would like to thank Professors P. S. Letelier (University of Campinas), W. A. Rodriguez Jr. (Unisal) and H. -J. He (University of Michigan) for very usefull comments. I am also in debt to Centro Universit\'ario Salesiano Unisal for the finantial support.

\section{References}
.
\par [1] P. A. M. Dirac, Proc. Roy. Soc. \textbf{A133} (1931) 60

[2] P. A. M. Dirac, Phys. Rev. \textbf{74} (1948) 817

[3] F. Rohrlich, Phys. Rev. \textbf{150} (1966) 1104

[4] D.Rosenbaum, Phys. Rev. \textbf{147} (1966) 891

\par [5] G. t 'Hooft, Nucl. Phys.\textbf{\ B79} (1974) 276;

\par A. M. Polyakov, Sov. Phys. JETP Lett.\textbf{\ 20} (1974) 194

[6] T. T. Wu and C. N. Yang, Phys. Rev. \textbf{D12} (1975) 3845

[7] R. A. Brandt and J. R. Primack, Phys. Rev. \textbf{D15} (1977) 1175

[8] H. -J. He, Z. Qiu and C. -H. Tze, Zeits. Phys. \textbf{C65} (1995), 
\par \textbf{hep-ph/9402293}

[9] Y. Aharonov and D. Bohm, Phys. Rev. {\textbf{115}} (1959) 485

[10] V. F. Weisskopf, Phys. Rev. \textbf{56} (1939) 72

[11] R. P. Feynman, F. B. Morinigo and W. G. Wagner, 
\par \textit{Feynman
Lectures on Gravitation},
\par Brian Hatfield Ed. Addison-Wesley Pub. Company
(1995), Lecture 11.

\end{document}